\documentstyle[10pt,aaspp4,flushrt]{article}  
\def\etal{\it et al. \rm }
\begin{document}
 
\title{AGN Activity in Giant LSB Galaxies}

\author{J. Schombert}
\affil{Department of Physics, University of Oregon, Eugene, OR 97403;
js@abyss.uoregon.edu}

\begin{abstract}
A search of large, HI-rich disk galaxies finds a significantly higher
fraction of low luminosity AGN signatures compared to other late-type galaxies.
Approximately half of the galaxies selected in this sample have AGN-like
behavior in their cores, the rest have HII nuclei resulting from simple
star formation.  Since AGN behavior is not evident in all the sample
galaxies, which where selected by high gas mass, we speculate that it is
the fuel flow rate that is the common feature between late-type LSB disks
and other active nuclear galaxies.

\end{abstract}


\section{INTRODUCTION}

There are several key questions to the evolution of active galactic nuclei
(AGN) phenomenon only a few of which have begun to be addressed with
current telescope technology. One concerns the amount of activity in the
cores of galaxies, in the sense of whether it is common for galaxies to
harbor a central engine or whether secondary effects, such as the
availability of gas for fuel, determines the existence and nature of an
AGN. A second problem as been the evolution of distant QSO's, in
particular, the fact that QSO's are so dominant in the past and, yet,
non-existent at the present epoch. It is assumed that the AGN in QSOs has
faded with time as their fuel supply was depleted. But, the low
luminosity, present-day counterparts have never successfully been
identified (Filippenko 1989).

The amount of energy released by AGN's occurs in such a small volume of
space, based on variability analysis, that the implication is that their
ultimate source of energy is gravitational release, most plausibly an
accretion disk surrounding a massive black hole.  An AGN is rarely
directly visible from the radiation emitted through a flat, nonstellar
photoionizing continuum; rather it is detected indirectly by reprocessed
energy through optical lines.  Therefore, although AGN's are a strong
source of $\gamma$-rays, x-rays and radio emission, detection by optical
emission lines is the most reliable method for classification at low
luminosities.

The occurrence of AGN's in normal and starburst galaxies has been the
subject of several studies (see Heckman 1987).  For example, Kennicutt,
Keel and Blaha (1989) find that half of galactic nuclei they surveyed
showed evidence of a secondary nuclear component with either AGN or LINER
characteristics.  Ho, Filippenko and Sargent (1997, hereafter HFS) found
that over half the galaxies in their magnitude-limited sample had AGN or
LINER nuclear spectra, most frequent in early-type spirals.  Many authors
claim that a majority of spiral nuclei are composite in nature, with a
central AGN surrounded by star forming regions (Pogge 1989).  This makes a
search for quesicent AGNs even more difficult since identification
requires the separation of star-forming emission features from true
non-thermal activity (i.e. the detection of low ionization lines
characteristic to Seyfert and LINER phenomenon).

Previous searches for `dead' AGN's have centered on deep spectra of the
cores of bright galaxies to detect low luminosity effects (Filippenko and
Sargent 1985, HFS) or dynamic surveys of nearby galaxies to detect the
high stellar velocities around massive black holes, assumingly the
fuel-starved engines of the past (see Kormendy and Richstone 1995).  This
study pursues a different sample set by selecting the class of giant, low
surface brightness (LSB) galaxies, known as Malin objects (Impey and
Bothun 1989). Large LSB galaxies have been shown to have an unusually high
occurrence of low level AGN activity, making them possible candidates for
the present-day remnants of QSO's (Sprayberry \etal 1993).  Malin 1 and
F568-6 (the prototypes to the Malin class) both display Seyfert 2
characteristics, although at very low luminosities (Impey and Bothun 1989,
Bothun \etal 1990).  Sprayberry \etal (1993) finds two of the eight large
LSB galaxies in their survey have broad line, AGN spectra in their cores.
This could be an observational selection effect, since it would be easier
to detect a low intensity line emission in objects with low surface
luminosity densities compared to the bright bulges of normal galaxies.
However, it is unlikely that deep spectral surveys to have missed similar
activity in higher surface brightness,i late-type galaxies (see HFS).

In order to determine if AGN activity is common in giant LSB galaxies, and
if such activity is unique to their intrinsic LSB nature, this project's
goal is to complete a spectroscopic survey of the nuclei of 34 giant, low
surface brightness and nine giant, high surface brightness (HSB) galaxies
selected to be massive in HI gas (M$_{HI} > 5\times10^{9}$ M$_{\odot}$)
and with diameters greater than 20 kpc.  In general, AGN host galaxies are
high in surface brightness with strong star formation signatures (Smith
\etal 1986, Heckman \etal 1986, HFS). The fact that giant LSB galaxies are
suspected to have a high rate of AGN activity, even at a low intensity
level, may suggest that AGN's are more frequent than previously thought
and that a connection exists between surface density (either gas or
stellar) and AGN intensity.

\section{OBSERVATIONS}

The data for this project was acquired on the Michigan-Dartmouth-M.I.T.
(MDM) 2.4m telescope located on the southwest ridge of Kitt Peak National
Observatory.  There were two runs in September 1992 and January 1993 using
the MkIIIb spectrograph at Cassegrain focus.  The spectrograph detector
was a TI CCD with 32 micron sized pixels.  The September run used a 300
lines mm$^{-1}$ grism blazed at 6000\AA\ with a resolution of 4.9\AA, the
January run used 300 lines mm$^{-1}$ grism blazed at 6400\AA\ with a
resolution of 6.6\AA.  A 2.8 arcsec slit was used to examined the 4800 to
8000\AA\ region of the spectrum, although there was no useful data below
5600\AA\ due to the poor blue response of the CCD.  The data was flattened
using turret flats.  Both runs suffered from light cirrus, thus
spectrophotometry calibration was not attempted.  All the sample galaxies
were exposed for 3600 secs between 1.0 and 1.2 of airmass.

Data reduction was performed using the IRAF spectra package.  Debiasing
was completed using overscan lines and dark frames.  Wavelength
calibration used Ne-Ar lamps taken before and after each object.  The
grism was centered on the rest frame velocity of H$\alpha$ emission.  Over
90\% of the sample displayed detectable optical emission.  Figure 1
displays several examples of emission near the H$\alpha$ line, rest frame
(marked by the arrow).  Also visible in each frame are the [N~II] feature
at 6583\AA\ and the [S~II] doublet at 6716 and 6731\AA.  Although the
wavelength range covers $H\beta$ and [O~III], these features are rarely
seen in the data due to poor blue sensitivity of the CCD and low S/N in
the emission lines themselves.  Of the final sample containing 45
galaxies, two were eliminated during reduction since the H$\alpha$ line
was buried in OH emission.

\section{DISCUSSION}

\subsection{Optical Emission}

The sample for in this project was drawn from the UGC (Nilson 1973)
with the intent of adopting a set of large, HI-rich disk galaxies over a
range of central surface brightnesses and between 5,000 and 15,000 km
sec$^{-1}$.  To this end, 34 LSB and nine high surface brightness (HSB)
galaxies were selected as listed in Table 1 along with their morphological
class, redshift, surface brightness category, HI mass, diameter and
presence of a sizable bulge (outside of morphological class).  The
division between HSB and LSB is roughly at central surface brightness of
22 $B$ mag arcsec$^{-2}$ or a mean surface brightness of 24 $B$ mag
arcsec$^{-2}$.  All the galaxies have diameters greater than 20 kpc and
M$_{HI} > 3.1\times10^{9}$ M$_{\odot}$ (the typical spiral is 15 kpc in
diameter and $1.5\times10^{9}$ M$_{\odot}$ in HI mass).  Figure 2 displays
the range of HI masses for the sample compared to the Huchtmeier and
Richter (1989) HI catalog of all galaxies types.  From this figure it can
be seen that the sample defines the extreme end of the HI mass function,
but this list is by no means complete, rather chosen to satisfy the
observing window.

The line emission detections and type of nuclear emission are also
summarized in Table 1.  Optical emission is, of course, common in disk
galaxies, usually associated with HII regions powered by the UV radiation
from young, massive stars.  Table 1 lists the characteristics of the
optical emission, whether the lines were detected in the nucleus as well
as disk, whether strong [N~II] or [S~II] was found and whether [O~I] was
detected.  The low-ionizations transitions, such as [N~II] at 6548,
6583\AA\ and [S~II] at 6716, 6731\AA\ are normally weak compared to
H$\alpha$ in objects whose spectra are dominated by HII regions.  Strong
[N~II] and [S~II] lines and the presence of [O~I] are taken to be an
indicator of an AGN behavior (see below).

Line emission is not always the best choice for AGN searches since optical
lines may be obscured by dust.  One advantage to searching a sample of LSB
galaxies is there has been no evidence for heavy dust obscuration based in
IRAS far-IR emission or deep imaging to detect absorbing lanes (Schombert
\etal 1990).  There is also very little current star formation in LSB
galaxies (McGaugh, Schombert and Bothun 1995) and what little does exist
is usually found in small, isolated HII regions in their outer disks
(McGaugh 1992).  Thus, there is the expectation that there will be little
contamination to nuclear emission lines due nearby star formation or
absorption from intervening dust clouds.  The nine HSB galaxies in this
study displayed evidence of dust in their disks, but their bulges appear
smooth and symmetric on blue Sky Survey prints.

The goal of this study is the separation of objects whose cores display
emission line radiation due to an exotic, non-stellar source (AGN) from
standard emission line radiation due to photoionization from star
formation (HII nuclei).  These non-stellar sources would include both
Seyfert and LINER (low ionization nuclear emission-line region) behavior,
although the nature of LINER activity is unclear (Filippenko 1996).  For
this study, we will define a core to be `active' if it has low ionization
features (strong [N~II] and [S~II]) combined with [O~I]. Thus, LINERS
would be included as active.  While it would be clearly preferable to have
several lines (such as [O~III]) for classification, the low S/N nature of
the blue region makes that impossible. Ho and Filippenko (1997) also found
that the ratios of H$\alpha$ to [N~II] and [S~II] were the primary
characteristic to distinguish HII nuclei from active ones. Thus, we will
adopt their definition of active versus HII nuclei and limit our
conclusions within this context of AGN-like behavior without determination
of the nature of this behavior to be Seyfert or LINER.

One clear example of the difference between HII regions and non-stellar
nuclear activity is the spectrum from the LSB galaxy UGC 4422.  A section
of the spectrum from the core and disk regions are shown in Figure 3.  The
core spectrum displays two of the most common features of active
classification in this study; strong [N~II] and [S~II] compared to
H$\alpha$ and [O~I] emission. Opposite characteristics are seen in the
disk spectrum, dominant and narrow H$\alpha$, no evidence of [O~I],
indicative of ionized gas from star formation.  While there is some weak
possibility that these features could be shock induced emission lines,
supernova remnants are unlikely to be undergoing shocks since cores of LSB
galaxies are low in gas surface density (McGaugh 1992).

\subsection{Nuclear Activity}

Of the 43 galaxies in this study, 41 (95\%) have detectable amounts of
H$\alpha$ emission in their core regions, coincident with the optical
center (see Table 1).  The spatial extend varys, although typically the
intensity of the emission was correlated to the underlying stellar
luminosity.  Of these 41 galaxies with nuclear emission, 21 (51\%) have
signatures of HII nuclei and 20 (49\%) have active signatures.  Galaxies
with active signatures (strong [N~II] and [S~II] plus [O~I] emission) all
displayed unresolved spatial extent as is expected for their nuclear
origin.  Although the data presented here was not calibrated, the
exposures were similar in depth with respect to distance.  A majority of
the 20 galaxies with an active classification have very weak emission
lines (based on comparison to continuum levels), such that the AGN would
be characterized as low luminosity (L(H$\alpha$) $< 10^{40}$ ergs
s$^{-1}$).  For comparison, HFS found from a magnitude-limited survey of
341 disk galaxies that 97\% of their sample had detectable amounts of
ionized gas in their cores and approximately 40\% of the nuclear emission
line galaxies could be regarded as active.  This is effectively identical
to the numbers found herein, despite radically different selection
criteria, i.e. selecting for HI mass and large metric diameter.

The two galaxies without any sign of nuclear or disk emission (UGC 11977
and 12740) are both LSB in nature.  Since the optical emission in LSB
galaxies is typically restricted to small isolated HII regions in the
outer disk, the lack of detection probably indicates a positioning of the
slit that fortuitously missed any HII regions, even in the core.  All the
HSB galaxies were detected in emission, usually with characteristic
rotation curves expected for bright, disk galaxies of various
inclinations.  We interpret the difference in detection rates as merely
reflecting the higher current star formation rates in HSB as compared to
LSB galaxies and the resulting enhanced frequency of bright HII regions.

The occurrence of core emission in the sample divides evenly by surface
brightness.  Of the HSB sample, nine of nine (100\%) have H$\alpha$
emission in their core regions, three (33\%) of which are classified as
active.  From the LSB sample, 32 of 34 galaxies (94\%) have H$\alpha$
emission in their core regions, 17 (53\%) of these are active.  The
fraction of emission nuclei is nearly identical between LSB and HSB
galaxies. Of that fraction, slightly more LSB galaxies are active, but not
significantly so.  We note that LSB and HSB galaxies have neither their
past star formation rates, their current star formation rates, their
morphological appearances, the disk kinematics, mass to light ratios, gas
ratios nor their molecular gas content in common.  Although the
evolutionary histories of LSB galaxies is unclear, it is obvious that
their development differs sharply from HSB galaxies.  Therefore, the
similar occurrence of AGN behavior between HSB and LSB galaxies is perhaps
the only characteristic that the two classes have in common and may
signify that the AGN mechanism is decoupled from the global disk
properties of a galaxy (see below).

In order to consider the effects of the central stellar population, the
sample was divided into those galaxies with morphologically distinct
bulges, based on visible inspection of POSS prints, and those without a
symmetric, central concentration of light.  Of the 43 sample galaxies, 32
were classified with bulges, 11 with none.  Of the 32 bulge galaxies, 31
(97\%) displayed nuclear H$\alpha$ emission, of which 17 (55\%) of these
had active signatures. Of the 11 galaxies without bulges, 10 of the 11
(91\%) have H$\alpha$ emission in their core regions, but only three
(27\%) had active signatures.  This is also in agreement with HFS who
found that very few spiral types without bulges have AGN signatures, even
though the number of HII nuclei was high among all spiral types.  Normally
one would expect a lower detection rate of emission lines against the
bright bulges of early-type spirals.  Thus, there is the expectation of a
higher detection rate in LSB galaxies is due to contrast enhancement.
However, the statistics herein suggest the opposite, reinforcing the
conclusion that AGN nuclei are more common in systems with bulges and that
the existence of a stellar population (and the associated gravitational
gradient) appears to be a prerequisite for an active nuclear region.

Although the initial comparison between this HI massive, large metric size
sample, and HFS's magnitude-limited sample indicates an identical
frequency of galaxies with active signatures (49\% for this sample versus
43\% for the entire HFS sample), this ignores the inherent basis
in this study to select late-type galaxies.  Of the original HFS sample,
341 of the 486 galaxies are disk systems (morphological class Sa or
later).  Of the disk galaxies, 331 displayed nuclear emission (97\%) of
which 132 (40\%) of these also displayed evidence of an active core.
Again, this is similar to the numbers found for this studies sample.
However, if the HFS sample is divided into early and late-type disk
galaxies (where early is defined from Sa to Sbc, late-type as Sc and
beyond) then there is a sharp difference in the frequency of active
nuclear signatures.  The early-type galaxies have 60\% of their nuclear
emission to be of the active variety, whereas the late-type subset has
only 17\%  with active nuclear cores.  Dividing this study's sample into
early and late-types (see Table 1) does not produce the same separation
with respect to active signatures.  Of early-type galaxies in this sample
(13 of the 43 galaxies), six have active signatures (46\%).  Among the
late-type galaxies (30 of the 43 galaxies), 14 active signatures (50\%),
identical within the counting statistics to the early-type galaxies.  The
most immediate conclusion to be drawn from this comparison of early and
late-type galaxies is that the sample selected for this study (HI massive,
large metric size) has a much higher occurrence of AGN signatures in the
late-type galaxies relative to the magnitude-limited sample of HFS, i.e.
gas supply has an impact on the detection of an AGN by optical emission.

\subsection{Broad Line AGN's in LSB Galaxies}

The widths of the active galaxies in this sample are typically quite
narrow, less than 2,000 km sec$^{-2}$.  Only five of the 20 galaxies with
active signatures have widths approaching 5,000 km/sec, which is
consistent with the estimate from HFS that their sample contained only
20\% of active systems with broad H$\alpha$ emission.

An interesting example a broad line LSB galaxy is UGC 6614, shown in
Figure 1.  UGC 6614 has similar characteristics to Malin 1 and F586-6
(Bothun \etal 1990 ), a large, HI-rich galaxies with large, low surface
brightness disk.  The appearence of the disk of UGC 6614 is also typical
of the Malin class of galaxies, weak, faint spiral arms with a distinct,
but small, bulge.  The H$\alpha$ line is broad, about 4,500 km sec$^{-1}$
FWHM with a lower peak intensity than the nearby [N~II] feature.  This
would lead to a Seyfert 2 classification, although on the low end of the
width distribution of AGN's, as is the case for Malin 1 and F568-6.

Three additional large LSB galaxies with AGN signatures were found by
Sprayberry \etal (1995), two with Seyfert 1 type lines of greater than
10,000 km sec$^{-1}$ and one with a Seyfert 2 width of 3,800 km
sec$^{-1}$.  From the small sample of this study and Sprayberry \etal, it
is difficult to draw any general conclusions other than to note that AGN
behavior in large LSB galaxies is not restricted to narrow line emission,
but most of the nuclei found in this study would be not be classed as Type
1.

\subsection{Fuel Deficiency Hypothesis}

There is no correlation between the classification of nuclear emission as
active and the gas mass of the underlying galaxy, although this is not
surprising since the entire sample herein represents the most extreme HI
mass objects in current galaxy catalogs, i.e. there is little range to
investigate the dependence on gas mass.  However, one trend is clear from
the data, in an HI-rich sample the frequency of AGN's for late-type LSB
galaxies is much higher than found in a magnitude-limited sample.

This correlation with LSB galaxies is contrary to studies that have shown
that most AGN host galaxies are high in luminosity and surface brightness
(Smith \etal 1986).  The only common trait among these bright, HSB host
galaxies, and the LSB Malin objects, is the gas fraction.  This leads one
to the speculation that it is not the stellar mass or the total mass of a
galaxy that is correlationed with the AGN behavior, but rather the gas
mass or the surface density.  And that the underlying physics is the
amount of gas available as fuel to the central engine.

Correlating the existence of AGN emission with gas properties has been
attempted by many studies.  In this instance, the occurrence and strength of
the optical emission is based on the availability and flow of the fuel supply.
Enhanced AGN emission within a particular class of galaxies can be due to any
number of processes that increases the flow of gas (for example, the
gravitational gradient associated with a bulge or a high surface gas density in
the inner disk) or simply a quesicent, static situation where the gas supply is
uninterrupted (for example, a lack of star formation in the core regions of LSB
galaxies).  The reverse can also be true where the fuel supply is cut off due
to ionization from the accretion disk in the low density gas.  The fuel
deficiency may explain why almost all disk galaxies either have an HII nuclei
or an AGN.

The higher frequency of AGN's in large LSB galaxies, those with an ample
fuel supply, would suggest that the fueling hypothesis has merit. In
addition, this hypothesis implies that there might be an AGN's present in
the core of all galaxies yet may remain quesicent because the fueling rate
is low (van der Marel 1997).  Since the intensity of the optical line
emission is also quite faint in LSB galaxies, this suggests that the
fueling rate is influenced by the local surface density of gas, not the
total gas mass.

\section{CONCLUSIONS}

This project has been a spectroscopic survey for low luminosity optical
nuclear emission in disk galaxies.  The sample used was selected to cover
a range of central surface brightness for the largest, and most HI-rich,
galaxies in the UGC catalog.  The suspicion is that large LSB galaxies
have a higher frequency of AGN behavior compared to HSB disk systems
(Sprayberry \etal 1995).  The results are summarized as the following:

\noindent{$\bullet$} Of galaxies selected by large absolute size and high
HI mass, 95\% have nuclear emission. Approximately 1/2 of those objects
display emission line characteristics indicative of non-thermal excitation
in the sense of having high ratios of [N~II]/H$\alpha$ and
[S~II]/H$\alpha$ and [O~I] emission, which is interpreted AGN-like
behavior.

\noindent{$\bullet$} The occurrence of AGN behavior was similar for the
high and low surface brightness members of the sample.  This is surprising
since the two families of galaxy types have little else in common, neither
their past star formation rates, current star formation rates,
morphological appearances, disk kinematics, mass to light ratios, gas
ratios nor molecular gas content.  This indicates that the AGN mechanism
is decoupled from the global disk properties of a galaxy.

\noindent{$\bullet$} The occurrence of AGN nuclei was highest in systems
with bulges, regardless of morphological type or mean surface brightness.
The existence of a stellar population (and the associated gravitational
gradient) appears to be a prerequisite for an active nuclear region.

\noindent{$\bullet$} The sharpest difference between this HI selected
sample and a magnitude selected sample was the frequency of AGN behavior
in late-type systems.  In a magnitude-limited sample (HFS), 60\%
early-type galaxies are found to have their nuclear emission of the active
variety, whereas the late-type subset had only 17\% with active nuclear
cores.  However, in this study, 46\% early-type galaxies have active
signatures whereas 50\% of the late-type galaxies are active, identical
within the counting statistics to the HI massive, early-type galaxies.
That there is no difference between the frequency of AGN's in the LSB
versus HSB subsets reinforces the fact that not all LSB galaxies are
late-type.  The LSB Malin objects have significant bulges and, therefore,
are mid to early-type in morphological appearance.

\noindent{$\bullet$} The excess number of galaxies with AGN signatures are
primarily LSB galaxies with bulges, the Malin objects.  The discovery of a
high fraction of AGN's in large LSB galaxies has an impact on AGN density
models since low-level AGN's could hide in a LSB galaxies population,
whose numbers are currently not known. The unknown number of giant LSB
galaxies, like Malin 1, opens the possibility that these objects are the
missing remnants of distant QSO's, although it is unclear how such a thin
disk could have survive an early QSO epoch.

Previous work (Smith \etal 1986) related AGN behavior with star formation
and high surface brightness.  The enhanced number of AGN's in LSB galaxies
indicates that another parameter other than star formation is correlated
with nuclear emission.  One candidate is, of course, the fuel supply which
is a source for both star formation and optical AGN emission.  Not all of
the HI massive galaxies studied here have AGN signatures, so gas mass is
clearly not primary.  Whereas the total fuel supply is necessary, it is
the amount of fuel available to the central engine that is more relevant.
Thus, it seems plausible that these galaxies share some common property
related to the fuel flow in the central regions.  This would also explain
the correlation between active nuclear emission and the presence of a
bulge, the bulge being a secondary consequence of a sharp gravitational
gradient that influences the inward fuel flow.  In any case, the future
study of large, LSB galaxies will find spectroscopic examination of the
central regions to be instructive.

\acknowledgements
I wish to thank MDM Observatories for the generous allocations of
telescope time needed to observe these difficult galaxies. Luis Ho
motivating me to dust off the data for publication, and I thank him for
his numerous papers on dwarf AGN activity.  And special thanks to the
referee, Rick Pogge, for an extremely helpful report and for pointing out the
importance of the [O~I] line information.   This research made use of the
NASA/IPAC Extragalactic Database (NED) which is operated by the Jet
Propulsion Laboratory, California Institute of Technology, under contract
with the National Aeronautics and Space Administration.  This research was
supported by NASA grant NAG5-6109 under the ADP program.

\figcaption{The above spectra display the H$\alpha$/[N~II]/[S~II] and
[O~I] features for six galaxies from the sample, all are classed as giant
LSB systems except for UGC 12511. The arrow signifies H$\alpha$ rest
frame.  Note the strong [N~II] and [S~II] relative to H$\alpha$, the
signature of an active core.}

\figcaption{A comparison histogram of the HI mass for the objects in this
study and the distribution from galaxies from the UGC catalog.  The
typical spiral is $1.5\times10^{9}$ M$_{\odot}$ in HI mass, all the
objects for this sample were selected to be greater than $5\times10^{9}$
M$_{\odot}$ in HI mass }

\figcaption{A comparison plot of emission from the disk and core of UGC
4422.  The core spectrum displays two of the most common features of
active classification in this study; strong [N~II] and [S~II] compared to
H$\alpha$, a broad H$\alpha$ width and [O~I] emission. The disk spectrum
displays dominant and narrow H$\alpha$ plus no indication of [O~I],
indicative of ionized gas from star formation.}

\clearpage
\begin{deluxetable}{llcrrrccccc}
\tablewidth{0pt}
\tablecolumns{10}
\small
\tablecaption{Spectroscopic Sample}
\tablehead{\colhead{Object} & \colhead{class} & \colhead{sfb} & \colhead{V} &
\colhead{$M_{HI}/M_{\sun}$} & \colhead{D (kpc)} & \colhead{disk em} &
\colhead{nuc em} & \colhead{OI?} & \colhead{active} & \colhead{bulge} }
\startdata
U369  & SA(rs)c    & H & 4366 &10.18 & 48.0 &Y &Y &N &N &N  \nl
U416  & SABdm      & L & 5346 & 9.93 & 20.2 &Y &Y &: &N &Y  \nl
U905  & S?         & L &11465 & 9.79 & 47.3 &N &Y &: &N &Y  \nl
U1230 & Sm:        & L & 3836 & 9.70 & 27.7 &Y &Y &N &N &N  \nl
U1378 & (R)SB(rs)a:& L & 2940 &10.12 & 34.4 &N &Y &Y &Y &Y  \nl
U1455 & SAB(rs)bc  & L & 5123 & 9.97 & 49.3 &Y &Y &Y &Y &Y  \nl
U1491 & Sm:        & L &12187 & 9.95 & 37.7 &Y &Y &Y &Y &N  \nl
U1752 & SA(s)cd   &  L &17836 &10.64 & 92.0 &N &Y &Y &N &Y  \nl
U1920 & (R)SB(s)a &  L & 6196 & 9.98 & 32.0 &Y &Y &Y &Y &Y  \nl
U1922 & S?        &  L &10894 &10.26 & 78.7 &: &Y &Y &Y &Y  \nl
U2360 & SAd:      &  L &10562 &10.15 & 43.6 &Y &: &: &: &N  \nl
U2399 & SAB(s)cd  &  L & 8006 & 9.79 & 35.8 &Y &Y &N &N &N  \nl
U2965 & Sm        &  L & 7319 & 9.98 & 22.7 &Y &Y &: &N &N  \nl
U2975 & Im?       &  L &19937 &10.53 & 89.1 &Y &Y &N &N &N  \nl
U3059 & SAdm      &  L & 4811 &10.00 & 43.0 &Y &Y &: &Y &Y  \nl
U3968 & SB(r)c    &  L & 6780 & 9.98 & 32.6 &Y &Y &: &: &Y  \nl
U4219 & SA(rs)b   &  L &12433 &10.47 & 89.8 &Y &Y &Y &Y &Y  \nl
U4422 & SAB(rs)c  &  L & 4300 & 9.91 & 47.3 &Y &Y &Y &Y &Y  \nl
U4658 & SA(rs)c   &  H & 4382 & 9.92 & 33.2 &Y &Y &Y &Y &Y  \nl
U4985 & SAB(s)b   &  L &10180 &10.26 & 56.0 &Y &Y &: &N &Y  \nl
U5366 & SAB(rs)c  &  H & 5141 &10.02 & 40.7 &Y &Y &: &N &Y  \nl
U6614 & (R)SA(r)a &  L & 6358 &10.16 & 37.2 &N &Y &Y &Y &Y  \nl
U6754 & SA(rs)b   &  L & 7023 &10.13 & 72.5 &: &Y &N &N &Y  \nl
U6968 & Sc        &  L & 8232 &10.25 & 79.3 &N &Y &Y &Y &Y  \nl
U7357 & SAB(s)c   &  L & 6683 &10.20 & 36.8 &Y &Y &Y &N &Y  \nl
U9061 & SB(r)bc   &  H & 5442 &10.61 &101.1 &Y &Y &N &N &Y  \nl
U9322 & SAB(s)bc  &  H & 5098 & 9.22 & 15.8 &Y &Y &: &Y &Y  \nl
U9631 & SB(rs)b   &  H & 1924 &10.18 & 45.7 &Y &Y &N &N &Y  \nl
U9733 & Dwarf     &  L & 6697 &10.01 & 23.0 &N &Y &N &N &N  \nl
U9797 & SB(r)b    &  H & 3144 &10.29 & 43.2 &Y &Y &N &N &Y  \nl
U10104& SA(rs)bc  &  L & 9841 &10.37 & 91.4 &Y &Y &N &N &Y  \nl
U10891& SAB(r)bc  &  H & 1665 &10.04 & 35.5 &Y &Y &: &N &Y  \nl
U10908& S?        &  L & 5818 &10.08 & 20.0 &Y &Y &N &N &N  \nl
U11578& Sdm       &  L & 4602 & 9.98 & 23.7 &Y &Y &N &N &Y  \nl
U11754& SABcd     &  L & 4825 & 9.90 & 29.9 &Y &Y &: &Y &Y  \nl
U11977& SABdm     &  L &11284 &10.05 & 42.7 &Y &N &N &N &N  \nl
U12128& S?        &  L &19507 &10.41 & 67.1 &N &Y &Y &Y &Y  \nl
U12289& Sd        &  L &10160 &10.05 & 45.4 &Y &Y &Y &Y &Y  \nl
U12511& Scd:      &  H & 3556 &10.05 & 30.6 &Y &Y &Y &Y &Y  \nl
U12740& Scd?      &  L &10522 & 9.83 & 36.2 &Y &N &N &N &Y  \nl
U12742& SB(s)cd:  &  L & 5554 &10.08 & 38.2 &Y &Y &N &N &N  \nl
U12845& Sd        &  L & 4880 & 9.90 & 35.2 &Y &Y &: &Y &Y  \nl
U12895& Sd        &  L & 6752 & 9.64 & 18.6 &Y &Y &N &N &Y  \nl
\enddata
\end{deluxetable}

\pagestyle{empty}
\clearpage
\begin{figure}
\plotfiddle{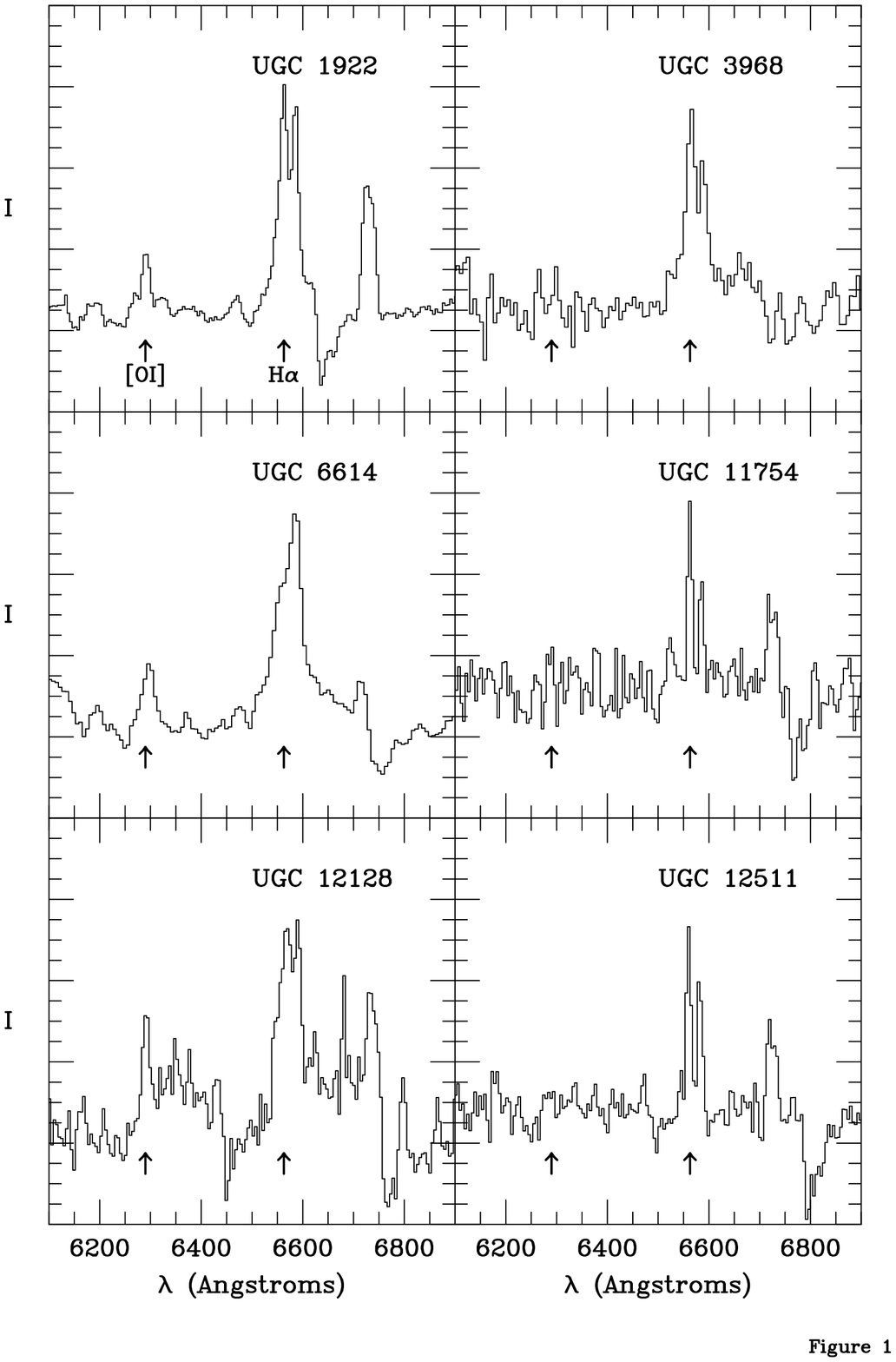}{11.5truein}{0}{100}{100}{-310}{170} \end{figure}

\clearpage
\begin{figure}
\plotfiddle{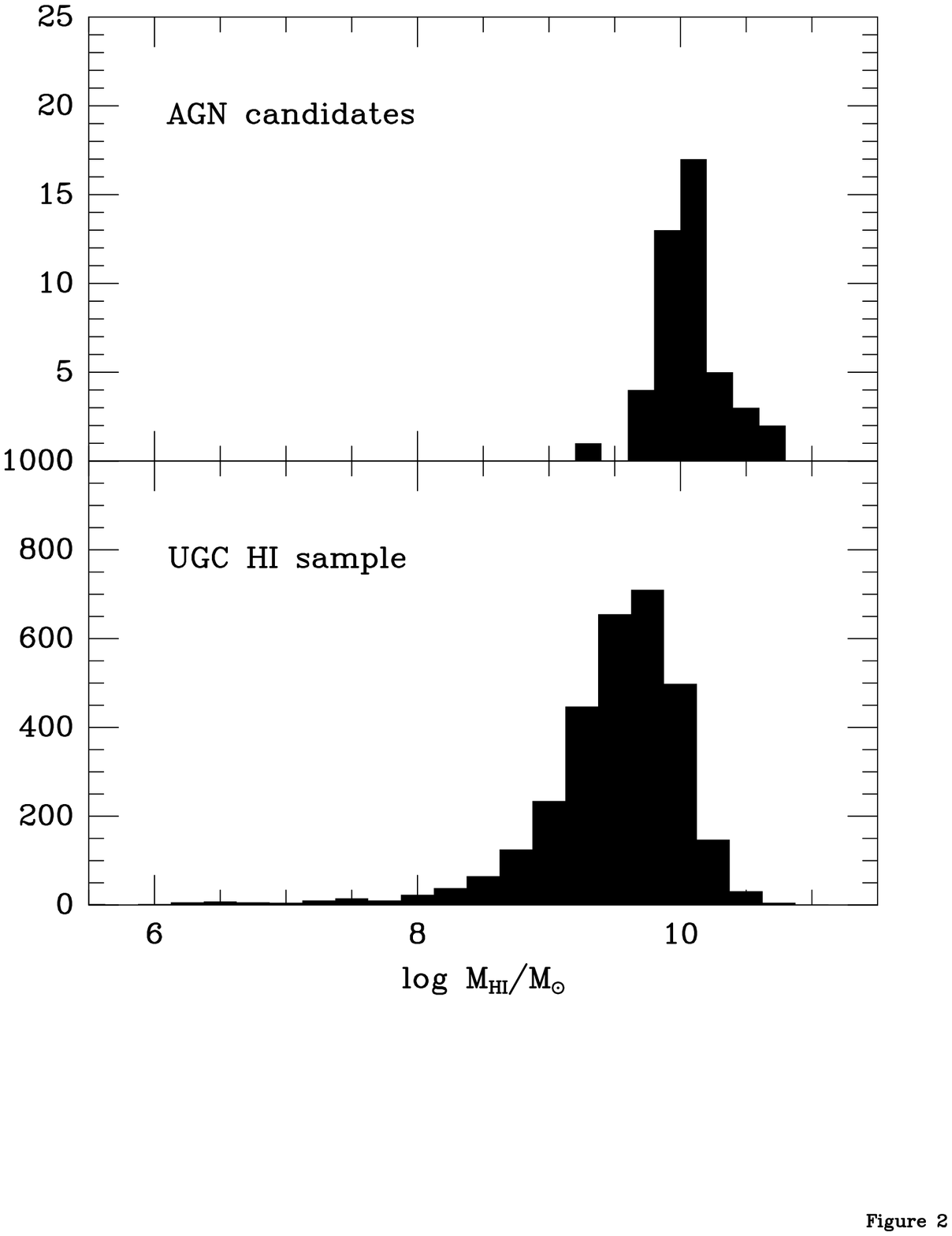}{11.5truein}{0}{100}{100}{-310}{170} \end{figure}

\clearpage
\begin{figure}
\plotfiddle{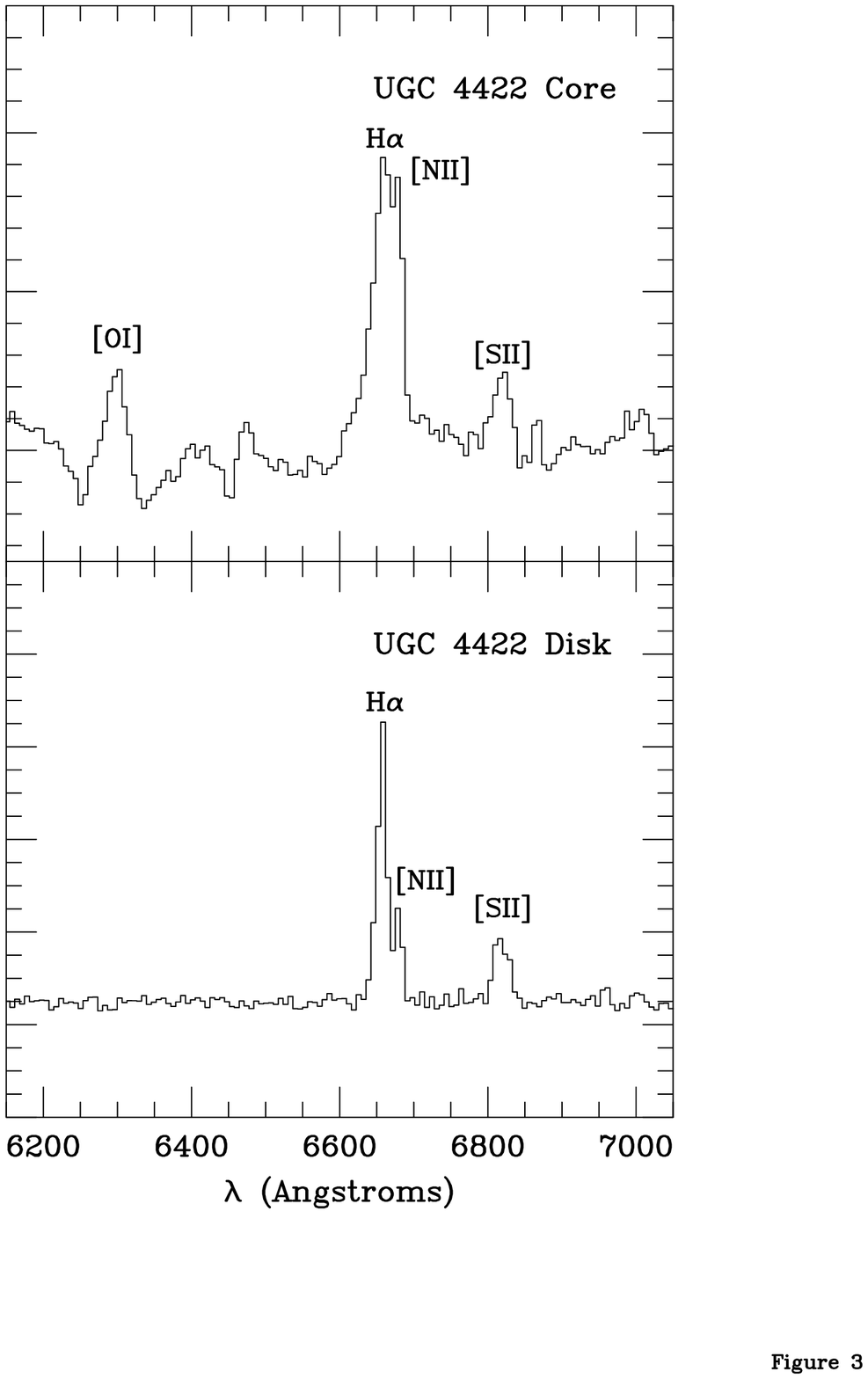}{11.5truein}{0}{100}{100}{-310}{170} \end{figure}

\end{document}